\documentstyle[twocolumn,art10]{article}
\begin{document}
\twocolumn[
\vspace*{8mm}
\mbox
\noindent
\Large Excitonic Instability and Origin of the Mid-Gap States \\
\newline
\normalsize Michael N. Kiselev \\
\newline
Department of Superconductivity and Solid State Physics,
Kurchatov Institute, 123182 Moscow, Russia \\
\newline
\hspace*{3ex}In the framework of the two-band model of a doped semiconductor
the self-consistent equations describing  the transition into
the excitonic insulator state are obtained for the 2D case.
It is found that due to the exciton-electron interactions the
excitonic phase may arise with doping in a semiconductor stable
initially with respect to excitonic transition in the absence of doping.
The effects of the strong interactions between electron (hole)
Fermi-liquid (FL) and excitonic subsystems can lead to the appearance of
the states lying in the middle of the insulating gap. \\ ]

\vspace{-10mm}
\par
Let us consider a two-band model of a doped semiconductor. On the one hand
this model allows us to describe the properties of a copper-oxide
superconductors if we have a $Cu-O$ model with bands formed by the hybridized
$Cu-d_{x^2-y^2}$ and $O-p_{x,y}$ orbitals [1,2].
On the other hand the two-band model has been used for studying the
transition from the semiconductor state into the Excitonic Insulator
State (EIS) [3,4]. In this case one considers that the Excitonic Transition
(ET)
arises only if the band gap $E_g$ is smaller than the binding energy $E_c$
of an exciton. The doping effectively decreases the exciton binding
energy and, therefore, it destroys  EIS.

\vspace*{-1mm}
The method based on the
three-body scattering approach to studying the influence of a
proximity to ET on the properties of the doping electron FL
has been proposed in [2]. The alternative method based on the solution of the
Bethe-Salpeter equation for electron-exciton scattering has been
developed in Ref.[4]. It was found there that if
$E_c < E_g <E_c+J$ (J is the binding energy of an exciton with an electron)
the creation of an exciton from the  vacuum would be favorable from the
energy standpoint. The exciton creation process will be stabilized by the
repulsion of excitons [3]. These arguments make it possible the
estimation for the equilibrium exciton density in 2D and 3D cases.
In the present work we intend to study the influence of an exciton
subsystem on the doping electron FL properties in the framework of the
method [4].

\vspace*{-1mm}
Following [3,4], let us consider the two-band model  of a doped
semiconductor
with the electron dispersion relation $\varepsilon_1$ and hole dispersion
relation $\varepsilon_2$ in 2D case.
$$\varepsilon_{i} = \pm E_g/2 \pm p^2/(2m_{i})
\hspace*{2ex}i=1,2  \eqno (1)$$
The Hamiltonian of interaction in terms of electron density operator
$\rho({\bf x})=\sum_{_{i}}\Psi^{\dagger}_{i}({\bf x}) \Psi_{i}({\bf x})$ is
$$ H_{int}= \frac{1}{2} \int d{\bf x} d{\bf y} \rho({\bf x})
\frac{e^2}{\mid {\bf x-y} \mid} \rho({\bf y}) \eqno (2)$$
Without any loss of generality we will consider the case of  doping into
the upper band and $m_2 >m_1$. The path integral  representation  of
partition function $Z$ of the system in terms of
slow electron field $\chi$ and excitonic field $\Phi$
describing the collective properties of electrons and holes [4]
has the form:
$$Z=\int\exp(S_{el}+S_{ex}+S_{int}) D \overline{\chi} D \chi
D \Phi^{*} D \Phi \eqno (3)$$

\noindent $
\displaystyle
S_{el}=\int_{0}^{\beta} d \tau \int d{\bf x} \overline{\chi}
\lbrack - \partial_{\tau} +\nabla^{2}/(2m_{1})+\mu \rbrack
\chi-
\hfill\\
\newline
\hspace*{4ex}-\frac{1}{2} \int_{0}^{\beta} d \tau \int d{\bf x} d{\bf y}
\mid\chi_{_{{\bf x},\tau}}\mid^{2}V^{eff}_{{\bf x}-{\bf y}}
\mid\chi_{_{{\bf y},\tau}}\mid^{2} \hfill\\
\newline
S_{ex}=\int_{0}^{\beta} d \tau \int d{\bf x} \Phi^{*}
\lbrack - \partial_{\tau}- \lambda({\bf x})
-\frac{f}{2} \mid \Phi \mid^{2} \rbrack \Phi \hfill\\
\newline
S_{int}=-\gamma \int_{0}^{\beta} d \tau \int d{\bf x}
\mid\chi_{_{{\bf x}, \tau}}\mid^{2}\mid\Phi_{{\bf x},\tau}\mid^{2}
\hfill (4)$
where $M=m_1+m_2$ is the mass of exciton, $V^{eff}$ is the effective Coulomb
potential of doping electrons, $f$ and $\gamma$ are, respectively,
the exciton-exciton and exciton-electron constants.
Moreover, $\gamma$ corresponds to the exciton-electron attraction [4].
The quantity $\lambda ({\bf x})=E_g-E_c[1-A(p_Fa_B)^2]-\nabla^2/(2M)$ is the
exciton
dispersion relation, $a_B$ is the first Bohr radius of electron and
the other notations have a standard form. The expression for $\lambda$
incorporates the circumstance that for the doping densities $p_Fa_B \ll 1$
the exciton binding energy falls linearly with increasing the
electron density, $A \sim 1$.

\begin{figure}[htb]\vspace{20mm}
\hfill\parbox[b]{70mm} {\footnotesize
Fig.1. Self-energy of excitons (a) and  electrons (b)} \end{figure}
\vspace*{-3mm}
The presence (or absence) of the excitonic condensate with the equilibrium
density $n_0$ in the system at $T=0$ is connected with the
presence of time-independent spatially-homogeneous solution of classical
equation for the saddle-point trajectory of field $\Phi$:
$$(i\partial_{t}-\lambda({\bf x})-W
- F\mid \Phi_{{\bf x},t}\mid^{2}) \Phi_{{\bf x},t}=0 \eqno (5)$$
where $W$ is the self-energy Fig.1(a),
$F \approx 4 \pi /(M\ln[E_{c}/\mu_{ex})$
is the renormalized vertex for the exciton-exciton interaction, and
$\mu_{ex}=n_{0}F$ is the exciton chemical potential.
The condition $\mu_{ex} \ll E_c$ corresponds to the
low density exciton gas $n_0a_{B}^2 \ll 1$.

Let us consider $\Gamma$ as a result of solution the  Bethe-Salpeter
equation for electron-exciton scattering.
$$\Gamma (P)=\gamma[1-i \gamma \int\frac{d^{3}k}{(2\pi )^{3}}G(k)g(P-k)]^{-1}
 \eqno(6)$$
where $G$ and $g$ are the Green's functions of electrons and excitons,
respectively, $d^3k =d{\bf k}d\epsilon$. On the Fermi surface
the vertex $\Gamma$ is a function only of the
electron-exciton total energy. To take self-consistently into account
the mutual influence of electron and exciton subsystems one should
solve the system of equations (7) with $\Gamma$ defined by relation (6),
see Fig.1:

\noindent $
\displaystyle
\hspace*{3ex}- \lambda ({\bf k}=0)+i\int\frac{d^{3}k}{(2\pi )^{3}} \Gamma (k)
G(k)=n_0F \hfill \\
\newline
\hspace*{3ex} \Sigma (k) =n_0 \Gamma (k) \hfill (7)$

\noindent
where the Green's functions are defined self-consistently by

\noindent $
\displaystyle
G=[\epsilon -\xi_{\bf p} - (\Sigma (\epsilon,{\bf p})-
\Sigma (0,{\bf p}_{F})+ i\delta sgn \xi_{\bf p}]^{-1} \hfill $

%\vspace*{2mm}
\noindent $
\displaystyle
g=2E_0({\bf P})[\omega^{2}-E^2({\bf P})+i\delta]^{-1},
\hfill (8)$

%\vspace*{2mm}
\noindent
and the spectra are $E^2({\bf P})=E^{2}_{0}({\bf P})+2\mu_{ex}
E_0({\bf P})$, \\
$\xi_{\bf p}=({\bf p}^2-{\bf p}_{F}^{2})/(2m_1)$,
\hspace*{1ex} $E_0({\bf P})={\bf P}^2/(2M)$.

Analysis of (6) and (7) shows that the equilibrium exciton density $n_0$
exists under the conditions $E_g-E_c < J$,
$\varepsilon_F < \varepsilon_{F}^{max} =\alpha J$, $\alpha \sim 1$.
In this case $n_0 \sim m_1 \varepsilon_{F}^{2}/J$,
$ \mid \Sigma (0,{\bf p}_{F}) \mid \sim \varepsilon_F$ and the effective
electron mass on the Fermi-surface is
$m^{eff}/m_1 =(1-(\frac{\partial \Sigma}{\partial \epsilon})_{_{0,p_F}})/
(1+(\frac{\partial \Sigma}{\partial \varepsilon(p)})_{_{0,p_F}}) \gg 1$.

\begin{figure}[htb]\vspace{25mm}
\hfill\parbox[b]{75mm} {\footnotesize
Fig.2. Schematic change in the one-particle DOS due to the exciton-electron
interaction} \end{figure}
Let us consider the electron DOS
$$ \nu(\epsilon) = -\frac{2}{\pi} \int \frac{d {\bf p}}{(2\pi)^2}
Im G^R({\bf p}, \epsilon) \eqno (9)$$
The pinning of the chemical potential and the origin of states
in the insulating gap (see Fig.2) are connected with the creation of the
complex
bound state of electron and excitons. The pseudo-gap that lies above the
Fermi level ($\delta \epsilon \sim J$) corresponds to the increase of $\Gamma$
(and $\Sigma$) near the threshold of the bound state formation.
For the region in which an exciton condensate exists, the strong interaction
between electron and exciton subsystems can lead to the marginal
properties of the electron FL in the normal state. In addition, there is a
possibility
of nonphonon superconductivity for this liquid. A further analysis of the
properties of such a system warrants a separate study.

This research was supported by the Russian Fund of Fundamental Research,
Grant No 93-02-02538.
\vspace*{-2mm}
\begin{flushleft}
{\small {\bf References}}
\end{flushleft}
\vspace*{-2mm}
{\footnotesize
1. C.M. Varma, S. Schmitt-Rink and E. Abrahams.
Solid State Communs., {\bf 62}, 681 (1987) \\
2. A.E.Ruckenstein and C.M.Varma. Physica C {\bf 185-189}, 134 (1991) \\
3. A.N.Kozlov and L.A.Maksimov. Sov. Phys. JETP {\bf 21}, 790 (1965),
L.V.Keldysh and A.N.Kozlov. Sov. Phys. JETP {\bf 27}, 521 (1968) \\
4. V.S.Babichenko and M.N.Kiselev. J. Mos. Phys. Soc., {\bf 2},
311 (1992), JETP Lett.{\bf 57}, 179 (1993), Physica C {\bf 209}, 133 (1993)}
\end{document}